\documentstyle[preprint,psfig,tighten,floats,aps]{revtex}

\def\ut#1{\rlap{\lower1ex\hbox{$\sim$}}{#1}}
\def\l{\ell_{P}}
\def\U{{\rm U}}
\def\SU{{\rm SU}}

\def\R{{\rm I}\!{\rm R}}

\def\i{\gamma}
\def\g{\gamma}
\def\j{{\vec j}}
\def\B{S}

\def\I{{\cal I}}
\def\P{{\cal P}}
\def\H{{\cal H}}
\def\k{\kappa}

\def\ket#1{{| #1 \rangle}}

\def\ba{\begin{eqnarray}}
\def\ea{\end{eqnarray}}
\def\be{\begin{equation}}
\def\ee{\end{equation}}

\preprint{\vbox{\baselineskip=12pt
\rightline{CGPG-98/4-2}}}

\begin{document}
\draft
\title{Quantum Geometry and Black Holes}
\author{Abhay Ashtekar\thanks{E-mail address: ashtekar@phys.psu.edu} 
and Kirill Krasnov\thanks{E-mail address: krasnov@phys.psu.edu}}
\address{Center for Gravitational Physics and Geometry,\\
Physics Department, Penn State, University Park, PA 16802, USA.}

\maketitle

\begin{abstract}

Non-perturbative quantum general relativity provides a possible
framework to analyze issues related to black hole thermodynamics from
a fundamental perspective. A pedagogical account of the recent
developments in this area is given.  The emphasis is on the conceptual
and structural issues rather than technical subtleties. The article is
addressed to post-graduate students and beginning researchers.

\end{abstract}

\section{Introduction}
\label{sec1}

In his Ph.D. thesis, Bekenstein suggested that, for a black hole in
equilibrium, a multiple of its surface gravity should be identified
with its temperature and a multiple of the area of its event horizon
should be identified with its thermodynamic entropy \cite{1}. In this
reasoning, he had to use not only general relativity but also quantum
mechanics. Indeed, without recourse to the Planck's constant, $\hbar$,
the identification is impossible because even the physical dimensions
do not match.  Around the same time, Bardeen, Carter and Hawking
derived laws governing the mechanics of black holes within classical
general relativity \cite{2}. These laws have a remarkable similarity
with the fundamental laws of thermodynamics.  However, the derivation
makes no reference to quantum mechanics at all and, within classical
general relativity, a relation between the two seems quite
implausible: since nothing can come out of black holes and since their
interiors are completely inaccessible to outside observers, it would
seem that, physically, they can only have zero temperature and
infinite entropy. Therefore the similarity was at first thought to be
purely mathematical. This viewpoint changed dramatically with
Hawking's discovery of black hole evaporation in the following year
\cite{3}.  Using an external potential approximation, in which the
gravitational field is treated classically but matter fields are
treated quantum mechanically, Hawking argued that black holes are not
black after all!  They radiate as if they are black bodies with a
temperature equal to $1/2\pi$ times the surface gravity. One can
therefore regard the similarity between the laws of black hole
mechanics and those of thermodynamics as reflecting physical reality
and argue that the entropy of a black hole is given by $1/4$-th its
area. Thus, Bekenstein's insights turned out to be essentially
correct (although the precise proportionality factors he had suggested
were modified).

This flurry of activity in the early to mid seventies provided
glimpses of a deep underlying structure. For, in this reasoning, not
only do the three pillars of fundamental physics  --quantum mechanics,
statistical physics and general relativity-- come together but
coherence of the overall theory seems to {\it require} that they come
together. It was at once recognized that black hole thermodynamics
--as the unified picture came to be called-- is a powerful hint for
the quantum theory of quantum gravity, whose necessity was recognized
already in the thirties.

Let us elaborate on this point. The logic of the above argument is
as follows: Hawking's calculations, based on {\it semi-classical}
gravity, lead to a precise formula for the temperature which can then
be combined with the laws of black hole mechanics, obtained entirely
in the {\it classical} framework, to obtain an expression of the black
hole entropy. Is there a more satisfactory treatment? Can one arrive
at the expression of entropy from a more fundamental, statistical
mechanical consideration, say by counting the number of
`micro-states' that underlie a large black hole?  For other physical
systems --such as a gas, a magnet, or the radiation field in a black
body-- to count the micro-states, one has to first identify the
elementary building blocks that make up the system. For a gas, these
are atoms; for a magnet, the electron spins; and, for radiation in a
black body, the photons. What then are the analogous building blocks
of a black hole? They {\it can not} be gravitons because the
gravitational fields under consideration are static. Therefore, these
elementary constituents must be essentially non-perturbative in
nature.  Thus, the challenges for candidate theories of quantum
gravity are to: i) isolate these constituents; ii) show that the number
of quantum states of these constituents which correspond to large
black holes in equilibrium (for which the semi-classical results can
be trusted) goes as the exponential of the area of the event horizon;
iii) account for the Hawking radiation in terms of quantum processes
involving these constituents and matter quanta; and, iv) derive the
laws of black hole thermodynamics from quantum statistical mechanics.

These are difficult tasks because the very first step --isolating the
relevant constituents-- requires new conceptual as well as
mathematical inputs. It is only recently, more than twenty years after
the initial flurry of activity, that detailed proposals have
emerged. One comes from string theory \cite{StringBH}
where the relevant elementary constituents are certain objects
(D-branes) of the non-perturbative sector of the theory.
The purpose of this article is to summarize the situation in another
approach, which emphasizes the quantum nature of geometry using
non-perturbative techniques from the very beginning. Our elementary
constituents are the quantum excitations of geometry itself and the
Hawking process now corresponds to the conversion of the quanta of
geometry to quanta of matter.

Although the two approaches are strikingly different from one another,
they are also complementary. For example, as we will see, we will
provide only an effective quantum description of black holes (in the
sense that we first isolate the sector of the theory corresponding to
isolated black holes and then quantize it). The stringy description,
by contrast is intended to be fundamental. Also, in our approach,
there is a one-parameter quantization ambiguity which has the effect
that the entropy and the temperature are determined only up to a
multiplicative constant which can not be fixed without an additional,
semi-classical input.  String theory, by contrast, provides the
numerical coefficients unambiguously. On the other hand, so far, most
of the detailed work in string theory has been focussed on extremal or
near extremal black holes which are mathematically very interesting
but astrophysically irrelevant. In particular, a direct, detailed
treatment of the Schwarzschild black hole is not available. Our
approach, by contrast, is not tied in any way to near-extremal
situations and can in particular handle the Schwarzschild case
easily. Finally, in the stringy approach, since most actual
calculations --such as derivation of the Hawking radiation-- are
carried out in flat space, the relation to the curved black hole
geometry is rather unclear.  In our approach, one deals directly with
the curved black hole geometry.

The article is organized as follows. Section 2 is devoted to
preliminaries.  We will first introduce the reader to basic properties
of quantum geometry in non-perturbative quantum gravity and then
briefly summarize a mathematical model called Chern-Simons theory which
is needed in the description of the quantum geometry of the horizon.
Main results are contained in section 3. We begin by describing the
classical sector of the theory corresponding to black holes and then
present the quantum description. By counting the relevant micro-states
we are led to the expression of entropy. Finally, we indicate how
Hawking evaporation can be regarded as a physical process in which the
quanta of the horizon area are converted to quanta of matter. As
suggested by the Editors, we shall try to communicate the main ideas
at a level suitable for a beginning researcher who is not already
familiar with the field. Therefore, the technicalities will be kept to
a minimum; details can be found in \cite{ABCK,BHRad,ack,abk} and
references cited therein.

\section{Preliminaries}
\subsection{Quantum Theory of Geometry}
\label{sec2.1}

In Newtonian physics and special relativity, space-time geometry is
regarded as an inert and unchanging backdrop on which particles and
fields evolve. This view underwent a dramatic revision in general
relativity. Now, geometry encodes gravity thereby becoming a dynamical
entity with physical degrees of freedom and Einstein's equations tell
us that it is on the same footing as matter.  Now, the physics of this
century has shown us that matter has constituents and the
3-dimensional objects we perceive as solids are in fact made of
atoms. The continuum description of matter is an approximation which
succeeds brilliantly in the macroscopic regime but fails hopelessly at
the atomic scale. It is therefore natural to ask: Is the same true of
geometry? Is the continuum picture of space-time only a
`coarse-grained' approximation which would break down at the Planck
scale?  Does geometry have constituents at this scale? If so, what are
its atoms?  Its elementary excitations? In other words, is geometry
quantized?

To probe such issues, it is natural to look for hints in the
procedures that have been successful in describing matter. Let us
begin by asking what we mean by quantization of physical quantities.
Take a simple example --the hydrogen atom. In this case, the answer is
clear: while the basic observables  --energy and angular momentum--
take on a continuous range of values classically, in quantum mechanics
their eigenvalues are discrete; they are quantized. So, we can ask if
the same is true of geometry. Classical geometrical quantities such as
lengths, areas and volumes can take on continuous values on the phase
space of general relativity. Can one construct the corresponding
quantum operators? If so, are their eigenvalues discrete? In this
case, we would say that geometry is quantized and the precise
eigenvalues and eigenvectors of geometric operators would reveal its
detailed microscopic properties.

Thus, it is rather easy to pose the basic questions in a precise
fashion. Indeed, they could have been formulated seventy years ago
soon after the advent of quantum mechanics. To answer them, however,
is not so easy. For, in all of quantum physics, we are accustomed to
assuming that there is an underlying classical space-time. One now has
to literally `step outside' space-time and begin the analysis
afresh. To investigate if geometrical observables have discrete
eigenvalues, it is simply inappropriate to begin with a classical
space-time, where the values are necessarily continuous, and {\it
then} add quantum corrections to it%
\footnote{This would be analogous to analyzing the issue of whether
the energy levels of a harmonic oscillator are discrete by performing 
a perturbative analysis starting with a free particle.}.
One has to adopt an essentially non-perturbative approach to quantum
gravity. Put differently, to probe the nature of quantum geometry, one
should not begin by {\it assuming} the validity of the continuum
picture; the quantum theory itself has to tell us if the picture is
adequate and, if it is not, lead us to the correct microscopic model
of geometry.

Over the past decade, a non-perturbative approach has been developed
to probe the nature of quantum geometry and to address issues related
to the quantum dynamics of gravity. The strategy here is the opposite
of that followed in perturbative treatments: Rather than starting with
quantum matter on classical space-times, one first quantizes geometry
and then incorporates matter. This procedure is motivated by two
considerations. The first comes from general relativity in which some
of the simplest and most interesting physical systems --black holes
and gravitational waves-- consist of `pure geometry'. The second
comes from quantum field theory where the occurrence of ultraviolet
divergences suggests that it may be physically incorrect to quantize
matter assuming that space-time can be regarded as a smooth continuum
at arbitrarily small scales.

The main ideas underlying this approach can be summarized as follows.
One begins by reformulating general relativity as a dynamical theory
of connections, rather than metrics% 
\footnote{As is often the case, it was later realized that this
general idea is not as heretical as it seems at first. Indeed, already
in the forties both Einstein and Schr\"odinger had given such
reformulations. However, while they used the Levi-Civita connections,
the present approach is based on chiral spin connections. This shift
is essential to simplify the field equations and to bring out the
kinematical similarity with Yang-Mills theory which in turn is
essential for the present treatment of quantum geometry.}
\cite{aa}.  This shift of view does not change the theory classically
(although it suggests extensions of general relativity to situations
in which the metric may become degenerate). However, it makes the
kinematics of general relativity the same as that of $SU(2)$
Yang-Mills theory, thereby suggesting new non-perturbative routes to
quantization. Specifically, as in gauge theories, the configuration
variable of general relativity is now an $SU(2)$ connection $A^i_a$ on
a spatial 3-manifold and the canonically conjugate momentum $E^a_i$ is
analogous to the Yang-Mills `electric' field. However, physically, we
can now identify this electric field as a triad; it carries all the
information about spatial geometry. In quantum theory, it is
natural to use the gauge invariant Wilson loop functionals, ${\cal P}
\,\exp \oint_\alpha\, A$, i.e., the path ordered exponentials of the
connections around closed loops $\alpha$ as the basic objects
\cite{js,rs1}.  The resulting framework is often called `loop quantum
gravity'.

In quantum theory, states are suitable functions of the configuration
variables. In our case then, they should be suitable functions
$\Psi(A)$ of connections. In field theories, the problem of defining
`suitable', i.e., of singling out normalizable states, is generally
very difficult and involves intricate functional analysis. In our
case, our space-time has no background metric (or any other
field). This makes the problem more difficult because the standard
methods from Minkowskian field theories can not be taken
over. However, although it is surprising at first, it also makes the
problem easier. For, the structures now have to be invariant under the
entire diffeomorphism group --rather than just the Poincar\'e group--
and, since the invariance group is so large, the available choices are
greatly reduced. This advantage has been exploited very effectively by
various authors to develop a new functional calculus on the space of
connections. The integral calculus is then used to specify the
(kinematical) Hilbert space of states and, differential calculus, to
define physically interesting operators, including those corresponding
to geometrical observables mentioned above.

For our purposes, the final results can be summarized as follows.
Denote by $\g$ a graph in the 3-manifold under consideration with $E$
edges and $V$ vertices. (For readers familiar with lattice gauge
theory, $\g$ can be regarded as a `floating' lattice; `floating'
because it is not necessarily rectangular. Indeed, we don't even know
what `rectangular' means because there is no background metric.)  It
turns out that the total Hilbert space $\H$ can be decomposed into
orthogonal, {\it finite dimensional} subspaces $\H_{\g, \j}$, where
$\j \equiv \{j_1,\, ... \, ,j_E\}$ stands for an assignment of a
non-zero half integer (`spin')  --or, more precisely, a
non-trivial irreducible representation of $SU(2)$--  to each of the
edges \cite{baezspin}. To specify a state in $\H_{\g,\j}$, one only has
to fix an intertwiner at each vertex which maps the incoming
representations at that vertex to the outgoing ones.%
\footnote{Given such a specification, we can define a function $\Psi$
on the space of connections as follows. A connection $A$ assigns to
each edge $e_I$ an $SU(2)$ element $g_I$ via parallel
transport. Consider the $(2j_I+1)\times(2j_I+1)$ matrix $M_I$
associated with $g_I$ in the $j_I$-th representation. Thus, with each
edge we can now associate a matrix. The intertwiners `tie' the row and
column labels of these matrices $M_I$ appropriately to produce a
number. This is the value of $\Psi$ on the connection $A$.  Note that
$\H_{\g,\j}$ is finite dimensional simply because the space of
intertwiners compatible with any spin-assignment $\j$ on the edges of
a graph $\g$ is finite dimensional.}
(For a trivalent vertex, i.e., one at which precisely three edges
meet, this amounts to specifying a Clebsch-Gordon coefficient
associated with the $j$'s associated with the three edges.)  Each
resulting state is referred to as a {\it spin network} state
\cite{rsspin}. The availability of this decomposition of $\H$ in to
finite dimensional sub-spaces is a powerful technical simplification
since it effectively reduces many calculations in quantum gravity to
those involving simple spin systems.

Since the edges of a graph are one-dimensional, the resulting quantum
geometry is polymer-like and, at the Planck scale, the continuum
picture fails completely. It emerges only semi-classically. Recall
that a polymer, although one dimensional at a fundamental level,
exhibits properties of a three dimensional system when it is in a
sufficiently close-knit state. Similarly, a state in which the
fundamental one-dimensional excitations of quantum geometry are
densely packed in a sufficiently complex configuration can approximate
a three dimensional continuum geometry. Individual excitations are as
far removed from a classical geometry as an individual photon is from
a classical Maxwell field. Nonetheless, these elementary excitations
do have a direct physical interpretation. Each edge carrying a label
$j= \frac{1}{2}$ can be regarded as a flux line carrying an
`elementary area': In this state, each surface which intersects only
the given edge of $\g$ has an area proportional to $\l^2$, where $\l$
is the Planck length. Thus, given a state in $\H_{\g,\j}$, quantum
areas of surfaces are concentrated at points in which they intersect
the edges of $\g$. Similarly, volumes of regions are concentrated at
vertices of $\g$. The microscopic geometry is thus distributional in a
precise sense.

In the classical theory, it is the triads that encode all the
information about Riemannian geometry. The same is true in the quantum
theory. The duals, $\Sigma^i_{ab}:= \epsilon_{abc} E^c_i$, of triads
are 2-forms and one can show that, when smeared over two dimensional
surfaces, the corresponding operators are well-defined and
self-adjoint on the Hilbert space $\H$ \cite{al1}. Thus, in the
technical jargon, triads are operator-valued distributions in two
dimensions. Various geometric operators can be constructed rigorously
by regularizing the appropriate functions of these triad operators
\cite{al1,rs,al2,tl,loll}.  Each is a self-adjoint operator and one
can show that they all have {\it purely discrete} spectra. Thus,
geometry is quantized {\it in the same sense} that energy and angular
momentum are quantized in the hydrogen atom. Properties of these
geometric operators have been studied extensively. However, for our
purposes, it will suffice to focus only on certain properties of the
area operators. We will conclude this section with a brief discussion
of these properties.

The {\it complete} spectrum of the area operators is known. The minimum
eigenvalue is of course zero. However, the value of the next
eigenvalue --the `area-gap'-- is non-trivial; it depends on the
topology of the surface. Thus, it is interesting that quantum geometry
`knows' about topology. Although the spectrum is purely discrete, the
eigenvalues $a_S$ crowd rather rapidly: for large eigenvalues $a_S$,
the gap $\Delta a_S$ between $a_S$ and the next eigenvalue goes as
$\Delta a_S \le \exp (-\, \sqrt{a_S}/\l)$. It is because of this crowding
that one can hope to reach the correct continuum limit.  More
precisely, the detailed behavior of the spectrum is important to
recover the correct semi-classical physics. This may seem strange at
first: Because the Planck length $\l$ is so small, one might have
thought that even if the spacing between area eigenvalues were
uniform, say $\Delta a_S = \l^2$, one would recover the correct
semi-classical physics. Detailed analysis shows that is not the
case. With an uniform level spacing, for example, one would not
recover the Hawking spectrum even for a large black hole \cite{BM}. To
obtain even a qualitative agreement with the semi-classical results it
is necessary that the eigenvalues crowd and the specific exponential
crowding one finds in this approach is also sufficient
\cite{al1,Carlo&}. Finally, let us describe the spectrum. For our
purposes, it is sufficient to display only the eigenvalues that result
when edges of a spin network state intersect the surface $S$ under
consideration transversely. These are given by \cite{al1,rs,ka}:
\begin{equation}
A_S=8\pi\gamma\l^2\sum_p\sqrt{j_p(j_p+1)},
\label{qarea}
\end{equation}
where the sum is taken over all points $p$ where edges of a spin
network state intersect $S$, $j_p$ are spin that label the
intersecting edges, and $\gamma$ is an undetermined real number,
known as the Immirzi parameter. Thus, the eigenvalues have an
ambiguity of an overall multiplicative factor $\i$ which arises from
an inherent quantization ambiguity in loop quantum gravity. As
mentioned in the Introduction, to fix this ambiguity, one needs
additional input, e.g., from semi-classical physics.

\subsection{Chern-Simons Theory}
\label{2.2}

It turns out that our analysis of black hole thermodynamics will
require, as a technical ingredient, certain results from a
three-dimensional topological field theory known as the Chern-Simons
theory. In this sub-section we shall recall these results very
briefly.

In this theory, the only dynamical variable is a connection one-form
$A$, which for our purposes can be assumed to take values in the Lie
algebra of $\SU(2)$. The theory is `topological' in the sense that it
does not need or involve a background (or dynamical) metric. Fix an
oriented three manifold $\Delta$ (which in our application will be a
suitable portion of the black hole horizon), and consider connections
$A$ on an $\SU(2)$ bundle over it. The action of the theory is given by:
\begin{equation}\label{csaction}
S_{CS}={k\over 4\pi}\int_{\Delta} {\rm Tr}
\left(A\wedge dA + {2\over 3}A\wedge A\wedge A\right).
\end{equation}
where $\k$ is a coupling constant, also known as the `level' of the
theory, and the trace is taken in the fundamental representation of
$\SU(2)$. The field equations obtained from the variation of this
action simply say that the connection $A$ is flat on $\Delta$. Such
connections can still be non-trivial if the manifold $\Delta$ has a
non-trivial topology. An especially interesting case arises when
$\Delta$ has a topology $\B \times R$ where $\B$ is a two-manifold
with punctures, i.e., points removed.  In this case, holonomies can be
non-trivial around punctures; i.e., we can have a distributional
non-trivial curvature which `resides at the deleted points', even
though on points included in $\B$, the curvature vanishes
everywhere. We will encounter this situation in the next section.

There are several ways to quantize Chern-Simons theory. For our
purposes we will need some facts about its canonical quantization.
Here, one has to first cast the theory into the Hamiltonian
framework. This is achieved by carrying out the canonical (2+1)
decomposition of the action, assuming that $\Delta$ has the topology
of $\B\times\R$, where $\B$ is a two-dimensional manifold. The phase
space consists only of the pullback of $A$ on $\B$; unlike the Yang-Mills
theory and general relativity, there are no `electric fields'. Thus,
the components of the connection do not Poisson commute:
\begin{equation}
\left\{ A_a^i(x), A_b^j(y) \right\} = {4\pi\over k}\,{\varepsilon}_{ab}
K^{ij}{\delta}^2(x-y).
\end{equation}
where $K^{ij}$ is the Cartan-Killing metric on $su(2)$, and
$\epsilon_{ab}$ is the (metric-independent) Levi-Civita density on
$S$. As in all generally covariant theories, the Hamiltonian turns out
to be a constraint
\begin{equation}
F_{ab}=0,
\end{equation}
where $F_{ab}$ is the pull-back of the curvature of $A$ on $S$. 

One is often interested in the quantization of this phase space in the
situation in which the two-manifold $\B$ has punctures.  Then, as
mentioned above, even though $F_{ab}$ vanishes on $\B$, the holonomies
around these punctures can be non-zero; intuitively, one can now
regard the curvature as being `distributional, residing at the missing
points'. To quantize the system, therefore, one has to provide certain
additional information --quantum numbers-- at these punctures.  The
resulting Hilbert space of states turns out to be finite dimensional,
and the dimension depends on the quantum numbers labelling
punctures. (See, e.g., \cite{km}.)

\section{Application to Black Holes}
\label{sec3}

We are now ready to examine black hole thermodynamics from the
fundamental perspective of non-perturbative quantum gravity. We will
summarize the overall situation following a systematic approach
developed in \cite{ABCK,ack,abk}. (For earlier work, see, e.g.,
\cite{BH}.)

In general relativity, supergravity or indeed in any modern classical
theory of gravity, using the causal structure one can define what one
means by a black hole in full generality, without having to restrict
oneself to any symmetries. With this notion at hand, one can then
restrict attention to specific contexts, such as stationary or
axi-symmetric situations. In quantum theory, the situation is not as
satisfactory: in none of the approaches available today, does one have
an unambiguous notion of a general, quantum black hole. Therefore,
within each approach, a strategy is devised to circumvent this
problem.  The strategy of the present approach is the following: We
will first pick out the sector of the classical theory that contains
isolated black holes  --the analogs of `equilibrium states' in
thermodynamics--  and quantize that sector. This will provide an
effective description. While this line of attack is sufficient for the
analysis of black hole entropy and provides an avenue to understanding
the Hawking radiation and laws of black hole mechanics from the
perspective of quantum gravity, it is far from being fully
satisfactory.  For example, since we {\it begin} with a black hole
sector of the classical theory, in a certain sense, the phase of the
Hawking process in which the black hole has fully evaporated can not be
encompassed in this approach. Consequently, as it stands, the approach
is unsuitable for a comprehensive analysis of issues, such as
`information loss', related to the final stages of the evaporation
process. The questions it is best suited to address are of the type:
``Given that a space-time contains a black hole of a certain size, what
is the associated statistical mechanical entropy and what is the
spectrum of the Hawking radiation?''  In a more fundamental
approach, one would first construct the full theory of quantum gravity,
single out, among solutions to the quantum field equations, those
states which are to represent a quantum black hole, and analyze their
physical properties.

So far, the detailed analysis exists only for non-rotating black holes
possibly with electric and dilatonic charges.  However, it is expected
that the main features of the analysis will carry over also to the
rotating case.  For simplicity, in the main discussion, we will focus
on uncharged black holes and comment on the charged case at the end.

\subsection{Phase space}
\label{sec3.1}

To single out the appropriate sector of the theory representing
isolated black holes, we will consider asymptotically flat space-times
with an interior boundary (the horizon) on which the gravitational fields
satisfy suitable boundary conditions. These conditions should be
strong enough to capture the physical situation we have in mind but
also weak enough to allow a large number of space-times.  A detailed
discussion of the appropriate boundary conditions can be found in
\cite{ack}. Here, we will only present the underlying ideas.

\begin{figure}
\centerline{\hbox{
\psfig{figure=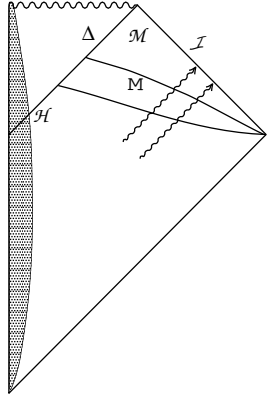,height=2.05in}}}
\caption{Space-times of interest.}
\end{figure}

To motivate our choice of the boundary conditions, let us consider the
space-time formed by a collapsing star (see Fig. 1). At some moment,
the event horizon $\cal H$ forms. A part of the gravitational
radiation emitted in the process falls in to the horizon along with
the matter and the rest escapes to infinity.  It is generally expected
that the black hole would finally settle down. Just as in the
calculation of entropy in thermodynamics one generally deals with
systems that have reached equilibrium, here we will be primarily
interested in the final phase in which the black hole has settled
down. We will make an idealization and assume that after a certain
retarded time no further matter or radiation falls in to the black
hole --the black hole is isolated-- and focus on the portion $\Delta$
of the event horizon to the future of this retarded time. Thus, in
this region, the area of any two-sphere cross-section of the horizon
will be constant, which we will denote by $A_H$.

Let us focus on the portion $\cal M$ of space-time, bounded by $\Delta$
and future null infinity ${\cal I}$, which can be regarded as the
space-time associated with the {\it isolated} black hole under
consideration.  For simplicity, we will assume that the vacuum
Einstein's equations hold on $\cal M$, i.e. that all the matter has
fallen in to the event horizon in the past of $\Delta$. Note that, as
the figure shows, there could still be gravitational waves in ${\cal
M}$ which escape to ${\cal I}$.  It is these space-time regions $\cal
M$ that constitute our sector of general relativity corresponding to
isolated black holes.

This sector can be isolated technically by specifying boundary
conditions which capture the intuitive picture spelled out above
\cite{ABCK,ack}.  It turns out, however, that, with our choice of
boundary conditions at $\Delta$, the standard general relativity
action in the connection variables is is no longer functionally
differentiable: The variation of the action with respect to the
connection contains a non-vanishing boundary term at
$\Delta$. However, it turns out to be possible to add to the action a
boundary term, whose variation exactly cancels the boundary term
arising in the variation of `bulk' action, thus making the total
action differentiable. Furthermore, the boundary term turns out to be
precisely the Chern-Simons action (\ref{csaction}), where the coupling
constant $k$ is now given by: $k = A_H/8\pi \gamma G$. It is quite
surprising that there is such a nice interplay between general
relativity, the boundary conditions for isolated black holes and the
Chern-Simons theory.

To canonically quantize the resulting theory, one has to cast it first
in the Hamiltonian form. This can be done in a standard fashion. As in
the case without black holes \cite{aa}, the basic canonical variables
are the pull-backs $A_a^i$ of the gravitational connection to a
spatial slice, such as $M$ in Fig. 1, and the triads $E^a_i$ on
$M$. However, these fields are now required not only to be
asymptotically flat but also to satisfy appropriate boundary
conditions on the two-sphere boundary $\B$ of $M$, i.e., the
intersection of $M$ and $\H$, which are induced by our boundary
conditions on space-time fields. First, the area of $\B$ defined by
the triads $E^a_i$ is a fixed constant, $A_H$. Second, the triads
and the connections are intertwined by the relation 
\be \label{bc}
\underline{F}_{ab}^i = - \frac{2\pi\gamma}{A_H}\,\, 
\underline{\Sigma}_{ab}^{i} \, ,
\ee
where, as before, $\Sigma_{ab}^i$ are the two-forms dual to the
triads, $F_{ab}^i$ is the curvature of the connection $A_a^i$ on $M$
and where the under-bars denote pull-backs to the two-sphere boundary
$\B$ of $M$. Finally, the boundary conditions also imply that the
pull-back of the connection $A_a^i$ to the boundary is reducible,
i.e., the (based) holonomies of $A_a^i$ along closed loops within $\B$
necessarily belong to an $U(1)$ sub-group of $SU(2)$.  Although it is
not necessary, for technical simplicity we can fix an internal vector
field, $r^i$, on $\B$ along which the curvature of the pulled-back
connection is restricted to lie, and partially gauge fix the
system. Then the gauge group on the boundary reduces from $\SU(2)$ to
$\U(1)$.  Furthermore, now only the $r$ component of (\ref{bc}) is
nontrivial.

As one might expect, due to the addition of the surface term to the
action, the fundamental Poisson brackets are modified; the symplectic
structure also acquires a surface term. This new contribution is just
the symplectic structure of the Chern-Simons theory on $\Delta$, where
the coupling constant $k$ is now given by
\begin{equation}\label{level}
k = {A_H\over 8\pi\gamma G}.
\end{equation}
Note that, up to a numerical coefficient, $k$ is simply the area of
the horizon of black hole measured in the units of Planck area $\l^2$.

The phase space consisting of the pairs $(A, E)$ satisfying our
boundary conditions is infinite dimensional.  It is clear, however,
that not all the degrees of freedom described by fields $A,E$ are
relevant to the problem of black hole entropy.  In particular, there
are `volume' degrees of freedom in the theory corresponding to
gravitational radiation propagating out to $\I$, which should not
be taken into account as genuine black hole degrees of freedom.  The
`surface' degrees of freedom describing the geometry of the horizon
$\B$ have a different status.  It has often been argued that the
degrees of freedom living on the horizon of black hole are those that
account for its entropy.  We take this viewpoint in our approach.

Note, however, that in the classical theory that we have described,
the volume and surface degrees of freedom cannot be separated: all
fields on $\B$ are determined by fields in the interior of $M$ by
continuity.  Thus, strictly speaking, classically there are no
independent surface degrees of freedom. However, as we described in
subsection \ref{sec2.1}, in the quantum theory the fields describing
geometry become distributional, and the fields on $\B$ are no longer
determined by fields in $M$; now there are independent degrees of
freedom `living' on the boundary. This striking difference arises
precisely because distributional configurations dominate in
non-perturbative treatments of field theories and would be lost in
heuristic treatments that deal only with smooth fields.  Furthermore,
as we will see in the next sub-section, it is precisely these quantum
mechanical surface degrees of freedom that account for the black hole
entropy!

\subsection{Quantization and Entropy}
\label{sec3.2}

To quantize the theory we proceed as follows. Since in the quantum
theory the volume and surface degrees of freedom become independent,
we can first quantize volume degrees of freedom by using the
well-established techniques of loop quantum gravity summarized in
section \ref{sec2.1}. We are interested only in those bulk states
which endow the boundary $\B$ with an area close to $A_H$. Boundary
conditions (\ref{bc}) then imply that, for each such bulk state, only
certain surface states are permissible. To calculate entropy, one
then has, in essence, just to count the states satisfying these
constraints.

\begin{figure}
\centerline{\hbox{
\psfig{figure=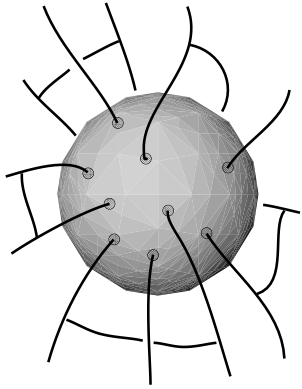,height=2in}}}
\bigskip
\caption{The flux lines of gravitational field pierce the black hole
horizon and excite curvature degrees of freedom on the surface. These
excitations are described by Chern-Simons theory and account for the
black hole entropy in our approach.}
\end{figure}

More precisely, the situation is as follows.  As discussed in
subsection \ref{sec2.1}, in the bulk, we have a `polymer geometry'
with one dimensional excitations. That is, a typical state is
associated with a graph $\g$ and the corresponding function $\Psi(A)$
of connections is sensitive only to what the connections do along the
edges of $\g$. Let us take one such state and denote by
$p_1,\ldots,p_n$ the intersection points of the edges of $\g$ with the
boundary $\B$ (see Fig. 2). For such a state to be admissible as a
micro-state of our large isolated black-hole --i.e., to belong to `the
quantum micro-canonical ensemble' underlying the classical black
hole-- two constraints have to be met at the boundary $\B$: i) the
quantum analog of the boundary condition (\ref{bc}) has to be
satisfied; and, ii) the area assigned by the state to $\B$ has to lie
in the range $(A_H -\l^2, \, A_H+\l^2)$.

As indicated in section \ref{sec3.1}, only the $r$ component of
(\ref{bc}) is non-trivial. Hence, in the quantum theory, our states
have to satisfy
\be \label{bc*}
\left(\hat{\underline{F}}_{ab}^ir_i + \frac{2\pi\gamma}{A_H}\, 
\hat{\underline\Sigma}_{ab}^i r_i \right)\,\cdot \Psi = 0.
\ee
Let us first consider $\underline{\hat{\Sigma}}$, the dual of the
triad operator. As indicated in section \ref{sec2.1}, this operator is
distributional and, on the boundary $\B$, the result of its action
turns out to be concentrated precisely at the points $p_1,\ldots,p_n$
where the edges of $\g$ intersect $\B$. Hence, the boundary condition
(\ref{bc*}) implies that in the quantum theory the state $\Psi$ has a
non-trivial dependence only on those connections $\underline{A}$ on
$\B$ which are flat everywhere except the points $p_1,\ldots,p_n$,
where its curvature is distributional. Recall that the Poisson
brackets between the connections $\underline{A}$ are precisely those
of the Chern-Simons theory. Thus the `surface part' of the state is
precisely a Chern-Simons quantum state associated with the two sphere
$\B$ with punctures $p_1,\ldots,p_n$. To summarize, the `volume part'
of the state encodes a polymer-type geometry in the bulk, while the
surface part is a Chern-Simons state, the two being `locked together'
by (\ref{bc*}).

Let us now turn to the second constraint on the states, namely the
requirement on area assigned to $\B$ by these states. To count the
number of states satisfying this constraint, let us examine the
eigenstates of the area operator which intersect $\B$, as above, in
$n$ points $p_1,\ldots,p_n$. These are precisely the spin network
states introduced in section \ref{sec2.1}. Therefore, each
intersection point $p$ now inherits a spin label $j_p$ from the edge
that intersects $\B$ in $p$. For notational simplicity, from now on,
we will refer to the pair $(p, j_p)$ as a puncture and often denote it
simply by $j_p$.  The area constraint tells us that we should restrict
our attention to the sets $\P$ of punctures,
$$\P=\{j_{p_1},\ldots,j_{p_n}\}$$ 
for which the area eigenvalue (associated to $\B$ by the corresponding
spin-network state)
$$a_\P = 8\pi\g\l^2 \sum_{p}\, \sqrt{j_p(j_p+1)}$$
lies in the interval $A_H -\l^2\le a_\P \le A_H+\l^2$. We will refer
to these $\P$ as the {\it permissible} sets of punctures.

Now, each set $\P$ of punctures gives rise to a Hilbert space of
Chern-Simons quantum states of the connection on $\B$. Denote it by
$\H_\P$. These are the surface states which are `compatible' with the
given permissible set $\P$ of punctures. Now, it follows
from standard results in Chern-Simons theory that, for a large number
of punctures, the dimension of ${\cal H}_\P$, goes as
\begin{equation}\label{dim}
{\rm dim} \H_\P \sim \prod_{j_p\in\P} (2j_p+1) ,
\end{equation}
Intuitively it is obvious that to calculate the entropy it suffices to
add up these dimensions, i.e. that the entropy $S_{\rm bh}$ is simply
\be\label{S}
S_{\rm bh} = \ln\,\, \sum_\P {\rm dim}\, (H_\P) \ee
where the sum extends over permissible sets of punctures.% 
\footnote{More formally, one can proceed as follows.  The total
Hilbert space carries information about {\it both}, the bulk and the
surface degrees of freedom.  We are not interested in this full space
since it includes, e.g., states of gravitational waves far away from
$\Delta$.  Rather, we wish to consider only states of the horizon of a
black hole with area $A_H$.  Thus we have to trace over the `volume'
states to construct an density matrix $\rho_{\rm bh}$ describing a
maximal-entropy mixture of surface states for which the area of the
horizon lies in the range $A_H \pm \l^2$.  The statistical mechanical
black hole entropy is then given by $S_{\rm bh} = - {\rm Tr} \rho_{\rm
bh} \ln \rho_{\rm bh}$.  As usual, this can be computed simply by
counting states, i.e., the right side of (\ref{S}).}
It is rather straightforward to calculate the sum for large $A_H$,
and we have:
\begin{eqnarray}\label{ent}
S_{\rm bh} = \frac{\i_0}{4\l^2\i}\,  A_H ,\quad {\rm where}\quad
\i_0 = \frac{\ln{2}}{\pi\sqrt{3}}.
\nonumber
\end{eqnarray}
(The appearance of $\i$ can be traced back directly to the formula for
the eigenvalues of the area operator, (\ref{qarea})).  Thus, in the
limit of large area, the entropy is proportional to the area of the
horizon.  If we fix the quantization ambiguity by setting the value of
the Immirzi parameter $\i$ to the numerical constant $\i_0$ (which is
of the order of $1$), then the statistical mechanical entropy is given
precisely by the Bekenstein-Hawking formula.

Are there independent checks on this preferred value of $\i$?  The
answer is in the affirmative.  One can carry out this calculation for
Reissner-Nordstrom as well as dilatonic black holes.  A priori it
could have happened that, to obtain the Bekenstein-Hawking value, one
would have to re-adjust the Immirzi parameter for each value of the
electric or dilatonic charge. This does {\em not} happen.  The entropy
is still given by (\ref{ent}) and hence by the Bekenstein-Hawking
value when $\i =\i_0$. The way in which the details work out shows
that this is quite a non-trivial check.

We conclude with two remarks.

1. An intuitive way to think about the quantum states that underlie an
isolated black hole is as follows. As indicated in section
\ref{sec2.1}, the edges of spin networks can be thought of as flux
lines carrying area. These flux lines endow the horizon $\B$ with its
area and also pin it (see Fig. 2), exciting the curvature degrees of
freedom at the punctures via (\ref{bc*}).  With each given
configuration of flux lines, there is a finite dimensional Hilbert
space ${\cal H}_\P$ describing the quantum states associated with these
curvature excitations. Heuristically, one can picture the horizon $\B$
as a pinned balloon and regard these surface degrees of freedom as
describing the `oscillatory modes' compatible with the pinning.

2. A detailed calculation of the black hole entropy shows that the
states that dominate the counting correspond to punctures all of which
have labels $j= 1/2$. For, in this configuration, the number of
punctures needed to approximate a given $A_H$ is the largest.  One can
thus visualize each micro-state as providing a yes-no decision or
`an elementary bit' of information.  It is very striking that this
picture coincides with the one advocated by John Wheeler in his ``It
from Bits'' scenario \cite{Wheeler}. Thus, in a certain sense, our
analysis can be regarded as concrete, detailed realization of those
qualitative ideas.

\subsection{Black Hole Radiance}
\label{sec3.3}

Since the number of micro-states of a large black hole grows as the
exponential of its area in Planck units, a quantum black hole is
system with an astonishingly large number of micro-states.% 
\footnote{ For example, a solar mass black hole has approximately
$\exp 10^{76}$ micro-states! Not only is this number extraordinarily
large compared to $ \exp 10^{23}$, the number we come across in
standard statistical systems, but it is very large even on
astronomical scales. For example, black body radiation with one solar
mass energy and at the same temperature as the sun has about 
$\exp 10^{63}$ micro-states.}
Hence, the statistical mechanical approximations one normally makes,
e.g., in treating the emission spectra in atomic physics, are
satisfied here very easily. Over twenty years ago, therefore,
Bekenstein \cite{B} used analogies from atomic physics to develop a
strategy for analyzing black hole radiance from microscopic
considerations. Most of the current work in this area uses this
strategy in one form or another. Our approach follows this trend.

The mechanism responsible for the radiance in our approach is based on
the following qualitative picture. Consider the micro-states of a
large black hole introduced in section \ref{sec3.2}, and let the black
hole be initially in an eigenstate $\ket{\Gamma}$ of the horizon area
operator. Radiation is emitted when the black hole jumps to a nearby
state $\ket{\Gamma'}$ with a slightly smaller area.  This change in
area corresponds to a change $\Delta M_{\Gamma\Gamma'}$ in the mass of
the black-hole which gets radiated away. Suppose, for simplicity, that
the emitted particle is of rest mass zero. Then, the frequency
$\omega_{\Gamma \Gamma'}$ of the particle when it reaches infinity is
given by $\Delta M_{\Gamma\Gamma'}=\hbar\omega_{\Gamma\Gamma'}$. Since
the area spectrum (\ref{qarea}) is purely discrete, the black-hole
spectrum is also discrete at a fundamental level: the emission lines
occur at specific discrete frequencies. At first one might worry
\cite{BM} that such a spectrum would be very different from the
continuous, black-body spectrum derived by Hawking using
semi-classical considerations. However, as indicated in
section \ref{sec2.1}, because the level spacing between the
eigenvalues of the area operator decreases exponentially for large
areas, the separation between the spectral lines can be so small that
the spectrum can be well-approximated by a continuous profile
\cite{Carlo&,al1}.

To determine the intensities of spectral lines and hence the form of
the emission spectrum, as in atomic physics, we can use Fermi's golden
rule. Thus, the probability of a transition $\Gamma \to \Gamma'$ with the
emission of a quantum of radiation is given by
\begin{equation}
W_{\Gamma\to\Gamma'}={2\pi\over\hbar}\,|V_{\Gamma\Gamma'}|^2\,
\delta(\omega-\omega_{\Gamma\Gamma'}){\omega^2 d\omega d\Omega\over
(2\pi\hbar)^3},
\label{fermi}
\end{equation}
where $V_{\Gamma\Gamma'}$ is the matrix element of the part of the
Hamiltonian of the system that is responsible for the transition,
$d\Omega$ is the element of the solid angle in the direction in which
the quantum is emitted, and $\omega_{\Gamma\Gamma'}$ is the frequency
of the quantum. (For brevity, we have suppressed the dependence of
this matrix element on initial and final states of the quantum field
describing the radiation.) The total energy $dI$ emitted by the system
per unit time in transitions of this particular type can be obtained
simply by multiplying this probability by $\hbar\omega$ times
the probability ${\rm Pr} (\Gamma)$ to find the system in the 
initial state $\ket{\Gamma}$:
\begin{equation}
dI(\Gamma \to \Gamma') = 2\pi \omega\,\, {\rm Pr} (\Gamma)\,\,
|V_{\Gamma\Gamma'}|^2\,\delta(\omega-\omega_{\Gamma\Gamma'}) 
{\omega^2 d\omega d\Omega\over (2\pi\hbar)^3}
\label{fermi-2}
\end{equation}

The question for us is: For quantum black holes, how do these
intensity distributions compare with those of a black-body? To answer
the question, one has to calculate the probability distribution ${\rm
Pr} (\Gamma)$ of the black hole, and the matrix elements
$|V_{\Gamma\Gamma'}|$ of the Hamiltonian responsible for
transitions. Recall, however, that, in atomic physics, the general
form of the spectrum is usually determined by the probability
distribution of initial states and by such qualitative aspects of the
underlying dynamics as selection rules for quantum transitions. The
detailed knowledge of the matrix elements is needed only to determine
the finer details of the spectrum. The situation for quantum black
holes is analogous.

The overall picture can be summarized as follows. For a large black
hole, since the typical energies of the emitted particles are
negligible compared to the black hole mass, it is reasonable to appeal
to basic principles of equilibrium statistical mechanics and conclude
that all accessible micro-states occur with equal probability. Now,
the entropy $S$ goes as the logarithm of the number of states. Hence,
the probability for occurrence of any one permissible micro-state
$\ket{\Gamma}$ is $\exp (-S)$. General considerations involving
Einstein's A and B coefficients now imply that the mean number
$n_{\omega lm}$ of quanta of frequency $\omega$ and angular momentum
quantum numbers $l,m$ emitted by the black hole per unit time is given
by
\begin{equation}\label{thermal}
n_{\omega lm}= {\sigma_{\omega lm}\over e^{\hbar\omega/T} - 1},
\end{equation}
where, $T$ is the thermodynamic temperature, 
\begin{equation} \label{temp}
{1\over T}: = {d S\over d M}
\end{equation}
and $\sigma_{\omega lm}$ is the absorption cross-section of the black
hole in the mode $\omega lm$. (See, e.g., \cite{B,BHRad}).

Although we have phrased the argument in terms of black holes --the system
now under consideration-- most of the reasoning used so far is quite
general. In the case of the black-hole, we know further that the
entropy $S$ is given by $A_H/4$, which fixes the thermodynamic
temperature in terms of the parameters of the black hole.  This turns
out to be precisely the temperature that Hawking was led to associate,
in the external potential approximation, with the spontaneously
emitted radiation. Recall that the initial semi-classical reasoning
had an unsatisfactory feature: the Hawking temperature is the property
of the emission spectrum at infinity and it is not a priori clear that
it is related to the thermodynamic properties of the black hole. The
reasoning given above, based on the micro-states of the quantum black
hole itself, fills this gap; the temperature in the emission spectrum
is indeed the statistical mechanical temperature of the black hole.

Finally, we can obtain the intensity spectrum. Since the s-waves
dominate in the emission from non-rotating black holes, we can focus
only on the $l=0$ modes. Then, to obtain the intensity spectrum, one
has to multiply $n_{\omega, l=0}$ by the energy $\hbar\omega$ and the
density of states $\omega^2 d\omega d\Omega$ in the mode, and integrate
out the angular dependence. The result is:
\begin{equation}\label{4}
I(\omega)d\omega \sim
{\hbar \omega^3 \sigma_\omega\, d\omega\over e^{\hbar\omega/T - 1}};
\end{equation}
Thus, very general arguments lead us to the result that the emission
spectrum of the black hole is thermal, in the sense that it has the
form (\ref{4}). The problem now reduces to that of calculating the
absorption cross-section $\sigma_\omega$ from first principles. Since
the Hawking analysis is semi-classical, there it was consistent to use
the classical value of this cross-section. In full theory, on the
other hand, we need to compute it quantum mechanically. Thus, the
remaining question is: Does the cross-section calculated in the full
quantum theory agree with the classical result? If so, one would have
a derivation of Hawking radiation from fundamental considerations.

Now, in the classical theory, the cross-section $\sigma_\omega$ in
(\ref{4}) is an approximately constant function of $\omega$ (equal to
the horizon area $A_H$).  In our quantum treatment, on the other hand,
there is an obvious potential problem. To see this, recall first from
section \ref{sec3.2} that the most likely micro-states are the ones in
which each puncture is labelled by $j=\frac{1}{2}$. Indeed they
dominate the distribution in the sense that they already provide the
leading order contribution to the entropy. Hence, one would expect
that, transitions $j=0 \to j=\frac{1}{2}$ would completely dominate
the absorption cross-section.  In this case, $\sigma_\omega$ would be
peaked extremely sharply at a single frequency and the final spectrum
(\ref{4}) would therefore look {\it very} different from Hawking's.
However, this transition is simply forbidden by a selection rule!
More precisely, if the interaction Hamiltonian responsible for the
transitions is gauge invariant --which it must be, on general physical
grounds-- its matrix element between the initial and final states of
the above type is identically zero. The situation here is again
analogous to that in atomic physics where the selection rules are
obtained simply by examining the transformation properties of the
interaction Hamiltonian under the rotation group.

\begin{figure}[t]
\centerline{\hbox{
\psfig{figure=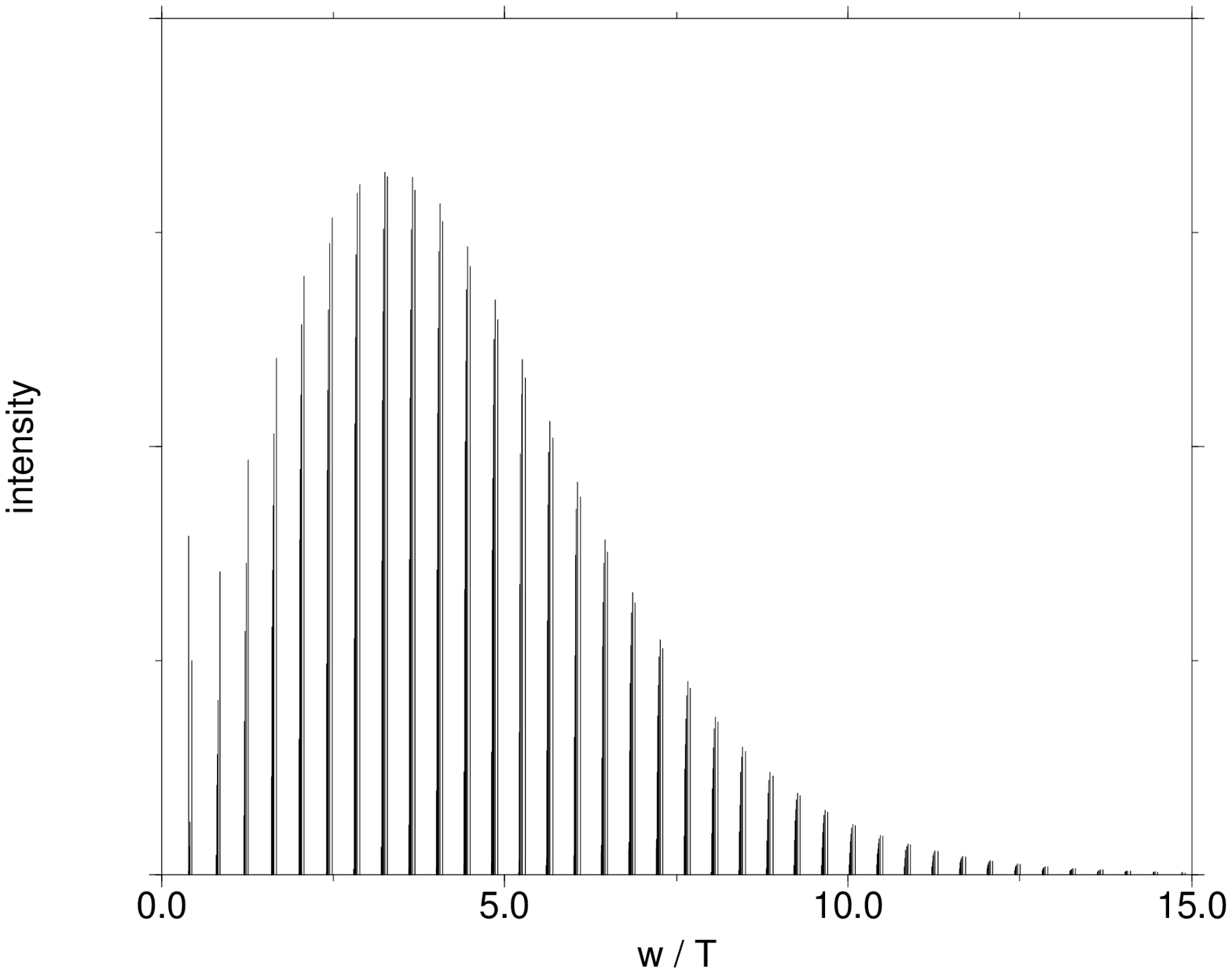,height=2in}}}
\centerline{\hbox{
\psfig{figure=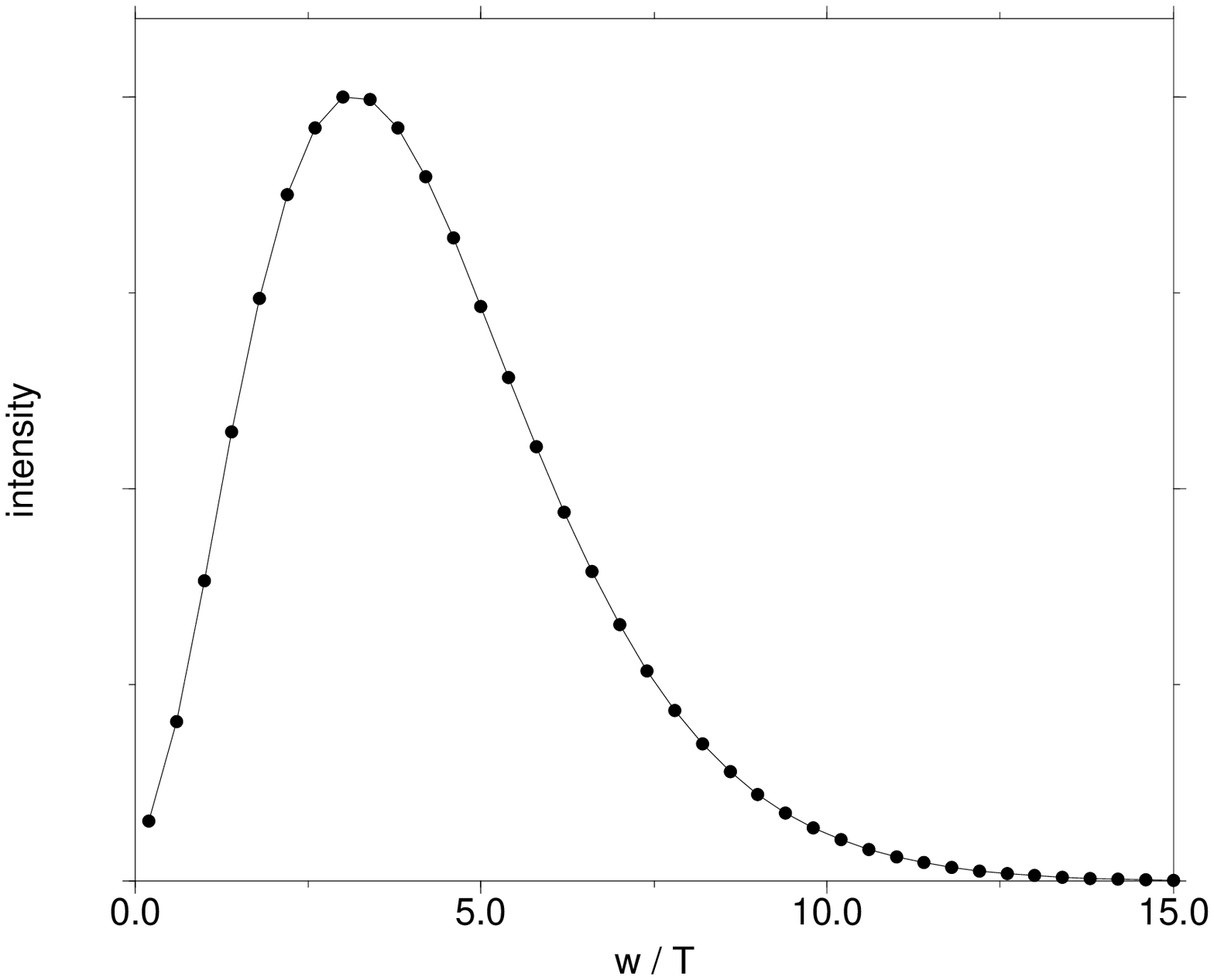,height=2in}}}
\caption{Emission lines with their intensities (above) 
and the total intensity of the radiation 
emitted per frequency interval as a function of frequency (below).}
\end{figure}

With this obstruction out of the way, one can proceed with the
calculation of the quantum absorption cross-section. Now one can argue
that, for a large class of Hamiltonians, the quantum absorption
cross-section is close to the classical one \cite{BHRad}. For
instance, if for allowed transitions the dependence on $\Gamma$ and
$\Gamma'$ of $V_{\Gamma, \Gamma'}$ is negligible, the intensity
spectrum is given by FIG 3. In the first plot, the intensities of
lines are computed using (\ref{fermi}). (Here, the overall
multiplicative factors are neglected; hence there are no units on the
$y$ axis.)  The second plot is obtained by first dividing the
frequency range into small intervals and then adding the intensities
of all the lines belonging to the same interval. It brings out the
fact that, although the emission occurs only at certain discrete
frequencies, the enveloping curve of the spectrum is thermal.

\section{Discussion}
\label{sec4}

The key ideas underlying our approach can be summarized as follows. In
the spirit of equilibrium statistical mechanics, we consider large,
isolated black holes (in four space-time dimensions). When appropriate
boundary conditions at the horizon are imposed, we find that the bulk
action of Einstein's theory has to be supplemented by a surface term
at the horizon which is precisely the Chern-Simons action.  We then
quantize the resulting sector of the theory. As one might expect from
the general structure of quantum field theories, the fields describing
geometry become distributional in the quantum theory.  Furthermore,
our background independent functional calculus tells us that the 
fundamental quantum excitations are of a specific type: they are
one-dimensional, like polymers.  The horizon acquires its area from
the points where these one-dimensional excitations pierce 
it transversely. Associated with each bulk state which endows the
horizon with a given area, there are `compatible' surface states on
the horizon itself which come from the quantization of Chern-Simons
theory thereon.  The total number of these `compatible' Chern-Simons
states grows exponentially with the area, whence the entropy is
proportional to the area of the horizon. Finally, using Fermi's golden
rule, one can compute the probabilities for transitions among these
horizon states.  Transitions that decrease the area are accompanied by
an emission of particles. From rather general considerations, one can
conclude that the emission spectrum is thermal, in agreement with
Hawking's semi-classical calculations.

At its core, this is a rather simple and attractive picture of quantum
black holes. However, it is far from being complete and further work
is being carried out in a number of directions. First, as in other
approaches, the calculations leading to the Hawking spectrum are based
on a number of simplifying assumptions. Although by and large these
are physically motivated, it is important to make sure that the final
results are largely insensitive to them. Second, all the detailed work
to date has been carried out only for non-rotating black
holes. Although the underlying ideas are robust, considerable
technical work will be required to extend the results to the
non-rotating case.%
\footnote{Curiously, as a general rule, the technical problems in
extending results from static situations to stationary ones turn out
to be `unreasonably difficult' in general relativity.  Examples that
readily come to mind are: the definitions of and theorems on multipole
moments, the discovery of black hole solutions, the black hole
uniqueness theorems, and, the analysis of stability under linearized
perturbations.}
Third, the role played by field equations in this program is yet to be
fully understood.  Since the construction of the phase space and the
associated symplectic structure is delicate, it is clear that, as it
stands, the framework is closely tied to general relativity; extension
to higher derivative theories, for example, will probably involve
significant modifications. (Extension to supergravity, by contrast, may
not.) Furthermore, in the quantum analysis, the `kinematic part' of
quantum Einstein's equations --the so called Gauss and diffeomorphism
constraints-- play a significant role. However, the role
of the Hamiltonian constraint is rather limited and needs to be better
understood. Finally, attempts to derive the laws of black hole
mechanics in full the quantum theory have just begun.

Perhaps the most puzzling and unsatisfactory aspect of the framework
is that there is an inherent quantization ambiguity which leads to a
one-parameter family of unitarily inequivalent theories. It shows up
in various technical expressions through the Immirzi parameter $\g$;
while the classical theory is completely insensitive to the value of
$\g$, the quantum theory is not. Roughly, this parameter is analogous
to the $\theta$ angle in Yang-Mills theory and as such its value can
be fixed only experimentally. Black hole evaporation can be taken to
be an appropriate experiment for this purpose. Now, it is
reasonable to assume that Hawking's semi-classical analysis \cite{3}
would agree with the experimental result for a 
large black hole. Hawking's answer is
recovered in our approach only when the Immirzi parameter is fixed to
be $\gamma_0 = \ln 2/\pi\sqrt{3}$. That is, it appears that, only for
this value of $\g$ would the non-perturbative quantum theory, with all
its polymer geometries and quantized areas, agree, in semi-classical
regimes, with the standard quantum field theory calculation in curved
space-times.  However, the issue is far from being settled. Perhaps a
new viewpoint will emerge and it may well cause the present picture to
change in certain respects. However, the picture does have a striking
coherence. Indeed, it is remarkable that results from three quite
different areas --classical general relativity, quantum geometry and
the Chern-Simons theory-- fit together without a mismatch to provide a
consistent and detailed description of the micro-states of black
holes. At several points in the analysis, the matching is delicate and
consistency could easily have failed. Since that does not happen, it
seems reasonable to expect that the lines of thought summarized here
will continue to serve as main ingredients also in the final picture.

\section{Acknowledgements} Much of this review is based on joint work with
John Baez and Alejandro Corichi. We are grateful to them and to the
participants of the 1997 quantum gravity workshop at the Erwin
Schr\"odinger Institute for many stimulating discussions. We are
especially grateful to John Baez for comments on early versions
of the manuscipt. The authors
were supported in part by the NSF grants PHY95-14240, INT97-22514 and
by the Eberly research funds of Penn State.  In addition, KK was
supported by the Braddock fellowship of Penn State.

\end{document}